\documentclass[a4paper,onecolumn,11pt]{IEEEtran}

\usepackage{epsfig,makeidx,color}
\usepackage{graphicx}
\usepackage{amsbsy}
\usepackage{amsmath}
\usepackage{amssymb}
\usepackage{euscript,verbatim}

\newcommand{\vol}{\mbox{vol}}

\newtheorem{prop}{Proposition}
\newtheorem{definition}{Definition}
\newtheorem{remark}{Remark}
\newtheorem{exam}{Example}
\newtheorem{thm}{Theorem}

\newcommand{\Z}{\mathbf{Z}} 
\newcommand{\R}{\mathbf{R}} 
\newcommand{\Q}{\mathbf{Q}} 
\newcommand{\C}{\mathbf{C}} 
\newcommand{\OO}{\mathcal{O}} 
\newcommand{\QQ}{\mathbb{Q}} 

\newcommand{\xv}{{\bf x}}

\newcommand{\Bc}{{\cal B}}

\newcommand{\Hc}{{\cal H}}

\begin{document}

\title{Nonasymptotic Probability Bounds for Fading Channels Exploiting Dedekind Zeta Functions}

\author{Camilla Hollanti, \emph{Member, IEEE}\thanks{C. Hollanti and D. Karpuk  are with the Department of Mathematics and System Analysis, P.O. Box 11100, FI-00076 Aalto University, Finland  (e-mails: camilla.hollanti@aalto.fi, davekarpuk@aalto.fi).}, Emanuele Viterbo, \emph{Fellow, IEEE}\thanks{E. Viterbo is with  the Monash University, Australia (e-mail: emanuele.viterbo@monash.edu).}, and David Karpuk, \emph{Member, IEEE} \thanks{The research of D. Karpuk and C. Hollanti was partly supported by the Emil Aaltonen
Foundation's Young Researcher's Project, and by the Academy of Finland 
grant \#131745.}\thanks{This research was partly carried while  C. Hollanti was visiting E. Viterbo at the Monash University in 2011.}\thanks{Part of this work was performed at the Monash Software Defined Telecommunications Lab
and was supported by the Monash Professional Fellowship and the Australian Research
Council under Discovery grants ARC DP 130100103.}\thanks{Part of the results  in Section \ref{bounds1} were presented at ICUMT 2011 \cite{ICUMT11}.}\thanks{AMS Classifications 14G50, 14G25.}
}


\maketitle

\begin{abstract}
In this paper, new probability bounds are derived for algebraic lattice codes. This is done by using the Dedekind zeta functions
of the algebraic number fields involved in the lattice constructions.
In particular, it is shown how to upper bound the error performance 
of a finite constellation on a Rayleigh fading channel and the probability of an eavesdropper's correct decision in a wiretap channel. As a byproduct, an estimate of the number of elements with a certain algebraic norm within a finite hyper-cube is derived. While this type of estimates have been, to some extent, considered in algebraic number theory before,  they are now brought into novel practice in the context of fading channel communications. Hence, the interest here is in small-dimensional lattices and finite constellations rather than in the asymptotic behavior.



\end{abstract}


\IEEEpeerreviewmaketitle

\section{Background and related work}

It has been well known for many years that number field lattice codes provide an efficient and robust means for many applications in wireless communications. We refer to \cite{OV} for a thorough introduction to the topic. More recently, wiretap channels and number field based codes have been under  study. Gaussian and fading wiretap channels have been considered for example in \cite{Hell_Gaussianwire,belfisoleoggis,belfisole,belfioggiswire,oggis_new}.  In \cite{BO_wiretap} the authors  propose to use number field lattice codes, which will also form the basis for our study and constructions. This paper can be seen, on one hand, as a continuation of \cite{ITW_camiame,ICUMT11}, where  analysis on lattice codes in fast and block fading channels was carried out based on various explicit code constructions and, on the other hand, of \cite{ISIT11-roopefransu,ITW11-roopefransu}, where Vehkalahti and Lu showed how
the unit group and diversity-multiplexing gain trade-off (DMT)
of division algebra-based space-time codes are linked to each
other through inverse determinant sums, and also demonstrated the connection to zeta functions and point counting. 
As continuation to \cite{ISIT11-roopefransu,ITW11-roopefransu}, the authors later showed that the density of unit group
completely determines the growth of the inverse determinant
sum, see  \cite{roopelaura_ISIT,roopefransujournal}.

Our work differs from \cite{roopefransujournal} in that it is targeted towards  practical SNR region and small delay rather than to asymptotic behavior. In addition to the pairwise error probability case, we apply the same methods to the wiretap channel, where an eavesdropper  is trying to intercept the data.   While in \cite{roopefransujournal} the authors concentrate on the number of units in a finite spherical subset of a lattice (as this is known to be the dominating factor), here we bound each individual term in the norm sum. This will enable us to estimate the number of points with each norm within a hyper-cubic constellation resulting in finer probability bounds. 


In summary, the contributions of this paper lie within
\begin{itemize} 

\item deriving (nonasymptotic) bounds for the probability expressions related to  Rayleigh fading channels and  to (Rayleigh fading) wiretap channels by using Dedekind zeta functions,
\item  finding more accurate bounds through geometric analysis based on the unit lattice, Dedekind zeta functions,    and bounded height norm sums,
\item deriving an estimate to the number of elements with certain norm  as a byproduct,  
\item demonstrating the accuracy of the estimate  through numerical examples and showing that the estimation error is very small for small dimensions.  
\end{itemize}

While bounded height (cf. Def. \ref{boundedheight})  estimates of the same type as the ones derived in this paper are known in algebraic number theory \cite{EV,EVLO}, they are far from being standard or well-known. Hence, we hope that our derivation and then the practical use of  such estimates will boost further research leading to yet tighter bounds. 

One should keep in mind that in general the estimation error is not negligible when the lattice dimension grows. However, as we apply these estimates in the situation where the lattice dimension corresponds to the decoding delay, the error is less severe. We will show that the estimate is very good when the dimension is relatively low and hence the delay short. Notice that the lattice dimension is not limiting the data rate as we can always increase the constellation size by choosing a bigger hyper-cube. Actually, the bigger the cube, the better our estimate will be, since the edge error effect becomes more negligible. Another limitation to the lattice dimension is forced upon by decoding -- it is known that the complexity of any ML decoder such as a sphere decoder grows exponentially in the lattice dimension.  We refer the reader to \cite{EV,EVLO} for more details regarding  norm forms and the treatment of  related error terms. We also point out the similarity of the constant term obtained here to the one in \cite{EVLO}, which well demonstrates the accuracy of our derivation. 

The rest of the paper is organized as follows. In Section \ref{preli}, we provide some algebraic preliminaries related to number field lattice codes. Section \ref{probinv} shortly introduces lattice coset coding employed in wiretap communications and revisits the probability expression for the fast  fading wiretap channel as well as for the typical Rayleigh fading channel, finally unifying the treatment of both expressions as one to simplify the computations in the rest of the paper. In Section \ref{bounds1}, first bounds are derived using Dedekind zeta functions. We then refine the bounds in Section \ref{geometric}, where geometric analysis is carried out in order to estimate the number of constellation points with certain algebraic norm. The accuracy of the estimate in low dimensions is demonstrated in Section \ref{examples}.
 Conclusions are provided in Section \ref{conclusions}.

\section{Algebraic preliminaries}
\label{preli}
 Lattices will play a key role throughout the paper, so let us first recall the notion of a lattice. For our purposes, a \emph{lattice} $\Lambda$ is a discrete abelian subgroup of a real vector space, 
$$
 \Lambda=\Z \beta_1\oplus \Z \beta_2 \cdots \oplus \Z \beta_t\subset \R^n,
$$
where the elements $\beta_1,\ldots, \beta_t\in\R^n$ are linearly independent, \emph{i.e.}, form a lattice basis, and $t\leq n$ is
called the \emph{rank} of the lattice. Here, we  consider full  ($t=n$) totally real lattices arising from algebraic number fields (see Def. \ref{can-emb} below). For more details on lattices and lattice codes, refer to \cite{OV}.

\begin{definition}\label{pdmin}
The \emph{minimum product distance} of a lattice $\Lambda$  is 
$$
d_{p,min}(\Lambda)=\min_{0\neq {\bf x}\in\Lambda}\prod_{i=1}^n|x_i|,
$$
where ${\bf x}=(x_1,\ldots,x_n)\in\Lambda.$ 
\end{definition}

\begin{definition} Let K be a number field. A \emph{real embedding} is a field homomorphism $\sigma:K\hookrightarrow \R$. A \emph{complex embedding} is a field homomorphism $\sigma:K\hookrightarrow \C$  such that $\sigma(K)\subsetneq \R$. 
\end{definition}
\begin{definition}\label{can-emb} Let $K/\Q$ be a totally real number field extension of degree $n$ and $\sigma_1,\ldots, \sigma_k$ its embeddings to $\R$. Let $\OO_K$ denote the ring of integers in $K$. The \emph{canonical embedding} $\psi:K\hookrightarrow \R^n$  defines a lattice $\Lambda=\psi(\OO_K)$ in $\R^n$:
$$
\psi(x)=(\sigma_1(x),\ldots,\sigma_k(x))\in\psi(\OO_K)\subset\R^n,
$$
where $x\in\OO_K$.  
\end{definition}
\begin{definition}
The \emph{algebraic norm} of $x\in K$ is defined as
$$
N_{K/\Q}(x)=\prod_{i=1}^n\sigma_i(x).
$$
We abbreviate $N(x)=N_{K/\Q}(x)$ whenever there is no danger of confusion.
\end{definition}

If $x\in\OO_K$, then $N_{K/\Q}(x)\in\Z$.
Hence, we have that
$$
d_{p,min}(\psi(\OO_K))=\min_{0\neq x\in\OO_K}|N_{K/\Q}(x)|=1.
$$

In what follows, a cubic constellation will be used, bounding the size of the vector components in the canonical embedding. To this end, we define the height of an algebraic integer as follows. 
\begin{definition}\label{boundedheight}
The \emph{height} of $x\in \OO_K $ is
$$
H(x)=\max_{1\leq i\leq n}|\sigma_i(x)|.
$$
We extend this notation to the height of a principal ideal $\mathcal{I}=(x),\, x\in \OO_K$, in a natural way by identifying an ideal $\mathcal{I}$ with its minimum-height generating element  and simply defining 
$$
H(\mathcal{I})=\min\{H(y)\,|\, \mathcal{I}=(y)\}.
$$ 
\end{definition}

Let us  denote by $(r_1,r_2)$ the \emph{signature} of $K$, \emph{i.e.}, $$[K:\Q]=n=r_1+2r_2,$$ where $r_1$ is the number of real embeddings $K\hookrightarrow \R$ and $r_2$ is the number of conjugate pairs of imaginary embeddings $K\hookrightarrow \C$. The group $\OO_K^\times$ of units of $\OO_K$ is described by the following well-known theorem, repeated here for the ease of reading.
\begin{thm}(\cite[Dirichlet Unit Theorem 1.9]{Nancy_CFT})\label{unitthm}
Let $K$ be a number field and let $(r_1,r_2)$ be the signature of $K$. There are units $\epsilon_1,\ldots,\epsilon_{r_1+r_2+1}\in\OO_K^\times$ such that
\begin{eqnarray*}
\OO_K^\times&\cong& \mathcal{W}_K\times \langle\epsilon_1\rangle\times\cdots\times \langle\epsilon_{r_1+r_2-1}\rangle\\
&\cong& \mathcal{W}_K\times \Z^{r_1+r_2-1},
\end{eqnarray*}
where $\mathcal{W}_K$ is the group of roots of units in $K$. The $\epsilon_j$ are called a \emph{fundamental system of units} for $K$.
\end{thm}

The fundamental units are used for defining the \emph{regulator} of $K$. Let $\{\epsilon_1,\ldots,\epsilon_r\}$ be a fundamental system of units for $K$, where $r=r_1+r_2-1$. Consider a matrix
$$
A=(\log|\sigma_j(\epsilon_i)|_j)
$$
for $1\leq i\leq r$ and $1\leq j\leq r_1+r_2$, and where we have used the notation
$$|x|_j=\left\{\begin{array}{ll}
|x| & \textrm{ if } 1\leq j\leq r_1,\\
|x|^2 & \textrm{ if } r_1\leq j\leq r_1+r_2.
\end{array}\right.$$
Here $|x|$ is the usual absolute value  on $\C$, and $\sigma_1,\ldots,\sigma_{r_1}$ are all the real embeddings,  while $\sigma_{r_1+1},\ldots,\sigma_{r_1+r_2}$ are a set of representatives of the imaginary embeddings.
\begin{definition} \label{regulator} The \emph{regulator} $\rho_K$ is the absolute value of the determinant of any $r\times r$ minor of $A$. It is independent of the choice of the fundamental system of units. 

The volume of the fundamental parallelotope of the log-lattice $\Lambda_{log}$ generated by $A$  is expressed in terms of the regulator as 
$$ \vol(\Lambda_{log})=\rho_K\sqrt{r_1+r_2}\,.$$
\end{definition}
In the case of a totally real number field we have 
$$
 \vol(\Lambda_{log})=\rho_K\sqrt{n}\,.
$$
 The regulator is a positive real number that in essence tells us how dense the units are. The smaller the regulator, the denser the units. Regulators can be easily computed by the Sage computer software \cite{sage}.

Let us conclude this section by defining the Dedekind zeta function and its truncated form. 

\begin{definition} \label{zeta}
The \emph{Dedekind zeta function} (cf. \cite[p. 37]{Nancy_CFT})  of a field $K$ is defined as
\begin{equation}\label{zeta-sum}\zeta_K(s)=\sum_{\mathcal{I}\subseteq\OO_K}\frac1{N_{K/\Q}(\mathcal{I})^s},\end{equation}
where $\mathcal{I}$ runs through the nonzero integral ideals of $\OO_K$. The sum converges for $\Re(s)>1$. Since $N_{K/\Q}(\OO_K)=1,$ we always have
\[
\zeta_K(s)>1
\]
for all integer exponents $s\geq 2$. From now on, we assume $2\leq s\in \Z$ since these are the  interesting values for the applications under study in this paper. Namely, $s=3$ corresponds to the wiretap case and $s=2$ to the PEP case. Values $s>3$ become relevant in the  multiple-input multiple-output (MIMO) case. Convergence is guaranteed in all these cases.

 The Dedekind zeta function can be written as a \emph{Dirichlet series} 
$$
\zeta_K(s)=\sum_{k\geq 1}\frac {a_k}{k^s},
$$
where $a_k=0$ for those $k$ that do not appear as a norm of an integral ideal.
\end{definition}

\begin{definition} \label{truncatedzeta}
The \emph{bounded height Dedekind zeta function}   of a field $K$ is defined as
\begin{equation}\label{zeta-sum}\zeta_K(s,m)=\sum_{0<H(\mathcal{I})\leq m}\frac1{N_{K/\Q}(\mathcal{I})^s}=\sum_{k=1}^{N_m}\frac {a_{k,m}}{k^s}\,, \end{equation}
where
$$
N_m=\max\{k\, |\, k=N(\mathcal{I}),\, H(\mathcal{I})\leq m\}.
$$
\end{definition}

Also the zeta functions as well as the Dirichlet coefficients $a_k$ can be computed by Sage \cite{sage}.

\section{Probability expressions and inverse norm sums}
\label{probinv}
\subsection{Wiretap channel and the probability of Eve's correct decision}
In a wiretap channel, Alice is transmitting confidential data to the intended receiver Bob over a fading channel, while  an eavesdropper Eve tries to intercept the data received over another fading channel. The security is based on the assumption that Bob's SNR is sufficiently large compared to Eve's SNR. In addition, a coset coding strategy \cite{Wyner} is employed in order to confuse Eve.

In coset coding, random bits are transmitted in addition to the data bits.
Let us denote the lattice intended to Bob by $\Lambda_b$, and by $\Lambda_e\subset\Lambda_b$ the sublattice embedding  the random bits. Now the transmitted codeword $\mathbf{x}$ is picked from a certain coset $\Lambda_e+\mathbf{c}$ belonging to the disjoint union 
$$\Lambda_b=\cup_{j=1}^{2^k} \Lambda_e+\mathbf{c_j}$$ embedding $k$ bits:
$$
\textbf{x}=\textbf{r}+\textbf{c}\in \Lambda_e+\textbf{c},
$$ 
where $\textbf{r}$ embeds the random bits, and $\textbf{c}$ contains the data bits.

We assume the fading is Rayleigh distributed and that both Bob and Eve have perfect channel state information (CSI-R), while Alice has none. The details of the channel model and related probability expressions  can be found in \cite{BO_wiretap}.

Next, let us recall the expression $P_{c,e}$ of the probability of a correct decision for Eve, when  observing a  lattice  $ \Lambda_e$.
 For the fast fading case \cite[Sec. III-A]{BO_wiretap},
\begin{equation}\label{fast-prob}
P_{c,e}\simeq \left(\frac 1{4\gamma_e^2}\right)^{n/2}\textrm{Vol}(\Lambda_b) \sum_{0\neq\mathbf{x}\in\Lambda_e}\prod_{i=1}^n\frac{1}{|x_i|^3},
\end{equation}
where $\gamma_e$ is the average SNR for Eve assumed sufficiently large so that Eve can perfectly decode $\Lambda_e$. This is a reasonable assumption, as Eve is assumed to have perfect CSI. Here $\Lambda_b$ denotes the lattice intended to Bob, and $\Lambda_e\subset \Lambda_b$. It  can be concluded from \eqref{fast-prob} that the smaller the sum
$$
\sum_{0\neq\mathbf{x}\in\Lambda_e}\prod_{i=1}^n\frac{1}{|x_i|^3},
$$
the more confusion  Eve is experiencing.

 As a construction method, the authors of \cite{BO_wiretap} propose to use the canonical embedding of the ring of integers $\OO_K$ (or a suitable proper ideal $\mathcal{I}\subset\OO_K$) of a totally real number field  $K$ over $\Q$. The field $K$ is chosen totally real to achieve full diversity. More precisely,  if $x\in\OO_K$, the transmitted lattice vector in the fast fading case would be 
\begin{equation}\label{embvector}
\mathbf{x}=\psi(x)=(\sigma_1(x),\sigma_2(x),\ldots,\sigma_{n}(x))\in\Lambda_e\subset \R^n,
\end{equation}
where $\psi$  denotes the canonical embedding (cf. Def. \ref{can-emb}) and $\sigma_i$ are the (now all real) embeddings of $K$ into $\R$. The corresponding probability of Eve's correct decision \eqref{fast-prob} yields the following \emph{inverse norm power sum}\footnote{The behavior of the probability of Eve's correct decision in some example cases has been analyzed in  \cite{ITW_camiame}.} to be minimized \cite[Sec. III-B]{BO_wiretap}:
\begin{equation}\label{fast-sum}
S_K=\sum_{0\neq x\in\OO_K}\frac{1}{|N_{K/\Q}(x)|^3}.
\end{equation}

\begin{remark}\label{infinite}
The above sum $S_K$ may not converge, since  infinitely many elements can have the same norm. This happens \emph{e.g.} when the unit group is infinite, which is the case for all field extensions other than the trivial one  and imaginary quadratic fields. In practice, however, we always consider finite signaling alphabets, so the sum becomes truncated and converges regardless of the field.
\end{remark}

Thus, we conclude by the following natural definition of the truncated version of $S_K$,
\begin{equation}\label{fast-sum-trunc}
S_K(m)=\sum_{0<H(x)\leq m}\frac{1}{|N_{K/\Q}(x)|^3}=\sum_{k=1}^{N_m}\frac {a_k}{k^3},
\end{equation}
where $N_m=\max\{k\,|\, k=N(x),\, H(x)\leq m\}$ is now defined for element norms  similarly as we did for ideal norms in the case of the truncated zeta function. 

Remember that $H(x)$ was only defined for algebraic integers, so the above notation implicitly contains the fact that $x\in\OO_K$.

\subsection{Rayleigh fading channel and the pairwise error probability}
Similar sums can be used to get  bounds for the  pairwise error probability  for the fast fading channel when employing the same number field code as in \eqref{embvector}. Let us use the  following Dirichlet series notation for the PEP (see \emph{e.g.} \cite{OV}), 
\begin{equation}\label{pep}
P_e=\frac{c}{\gamma^n}\sum_{0\neq x\in\OO_K}\frac{1}{|N_{K/\Q}(x)|^2}=\frac{c}{\gamma^n}\sum_{k\geq 1}\frac{c_k}{k^2},
\end{equation} 
where $\gamma$ is the SNR and $c$ is a fixed constant. We shortly denote the truncated version (ignoring constants) as
\begin{equation}
\label{pepsum}
S_{K,PEP}(m)=\sum_{0<H(x)\leq m}\frac{1}{|N_{K/\Q}(x)|^2}=\sum_{k=1}^{N_m}\frac{c_k'}{k^2}.
\end{equation}

\subsection{Unified treatment of the norm sums}
Let us use a more general notation 
\begin{equation}\label{fast-sum}
S_K(s)=\sum_{0\neq x\in\OO_K}\frac{1}{|N_{K/\Q}(x)|^s}.
\end{equation}

Now we can carry out the analysis for both the sums \eqref{fast-sum-trunc} and \eqref{pepsum} by simply denoting 
\begin{equation}\label{combinedsum}
S_K(s,m)=\sum_{0<H(x)\leq m}\frac{1}{|N_{K/\Q}(x)|^s}=\sum_{k=1}^{N_m}\frac {a_k}{k^s},
\end{equation}
and then inserting $s=2$ in the Rayleigh fading case, and $s=3$ in the wiretap case.

\section{First bounds}
\label{bounds1}
In this section, we will derive lower and upper bounds for the inverse norm power sum (cf. \eqref{fast-sum-trunc}, \eqref{pepsum}) by using the Dedekind zeta functions (cf. Def. \ref{zeta}). 

 Similarly to the zeta function, we trivially have that $S_K(s)>1$. Albeit straightforward, the following result gives us a nontrivial lower bound $\neq 1$ for the sum $S_K(s)$.

\begin{prop}(Lower Bound)

Assume that $\OO_K$ is a principal ideal domain (PID) and  $\Lambda_e$ is as above with $x\in\OO_K$. Let $m$ denote the  maximum  height included in the sum. The Dedekind zeta function  $\zeta_K(s)$ ($1<s\in\Z$) 
 provides us with a  lower bound for bounded the height sums $S_K(s,m)$, \emph{i.e.},
$$
S_{K}(s,m)>\zeta_K(s,m) > 1,
$$

If $K$ is the field or rationals or an imaginary quadratic number field, also the unbounded sum will converge.

\end{prop}
\begin{IEEEproof}
Note that $N_{K/\Q}(\mathcal{I})=[\OO_K:\mathcal{I}]\in\Z_+$, and that in the zeta function the summation only goes through the (integral) ideals of $\OO_K$, whereas in $S_K$ we sum over all the algebraic integers of $K$. 

Let us denote by  $$S_K=\sum_{n\geq 1}\frac{b_k}{k^{s}}\textrm{ \ and \ \  } \zeta_K(s)=\sum_{n\geq 1}\frac{a_k}{k^{s}}$$ the Dirichlet series  \cite[p. 31]{Nancy_CFT} of $S_K$ and $\zeta_K(s)$. Denote further by  
$$
A=\{k\,|\, a_k\neq 0\}\subseteq \Z_+
$$ 
the set of values $k$ that appear as norms in $\zeta_K(s)$, and by
$$
B=\left\{k\ |\, b_k\neq 0\right\}\subseteq \Z_{+}
$$ 
the set of values that appear as norms  in $S_K$.
 As $K$ is a PID, we know that $N_{K/\Q}(\mathcal{I})=\min_{0\neq x\in\mathcal{I}}|N_{K/\Q}(x)|$, and that $N_{K/\Q}(\mathcal{I})=N_{K/\Q}((\alpha))=|N_{K/\Q}(\alpha)|$, where $\alpha$ is the generator of $\mathcal{I}$. Hence, we have that 
$$
A=B.
$$
 Further, we easily see that $b_k\geq a_k$. Namely, if $k$ appears as a norm for distinct ideals $\mathcal{I}_i=(\alpha_i),\, i=1,\ldots,a_k$, it then appears as a norm at least for the (distinct) elements $\alpha_i\in\OO_K$. In addition, we may have an element $\alpha\in\OO_K,\,\alpha\neq \alpha_i\,\forall i$ with the same norm $n$. On the other hand, if $k$ appears as a norm for distinct elements $\alpha_i\in\OO_K,\,i=1,\ldots,b_k$, it cannot appear as a norm for an ideal for more than $a_k=b_k$ times, since $\OO_K$ is a PID. This proves the claimed lower bound for the infinite sums. Now it is obvious that the bound also holds for the bounded height sums, where the maximum height is the same for both sums.  

\end{IEEEproof}

\begin{remark}\label{whynottotreal}
According to Dirichlet's Unit Theorem (cf. Prop. \ref{unitthm}) the rank of the group of units in $\OO_K$ is $r=r_1+r_2-1,$ where $(r_1,r_2)$ is the signature of $K$, that is, $n=r_1+2r_2$. Therefore the infinite sum $S_K$ does not  converge for totally real number fields $\supsetneq \Q$ ($r_1>1$), as they have an infinite unit group (cf. Remark \ref{infinite}). Hence, in the present application case where $K$ is totally real, we indeed always need to consider truncated sums. In practice the constellations are always finite, hence this poses no real restrictions. 
\end{remark}

\begin{remark}  Dedekind zeta functions have also been used in \cite{ISIT11-roopefransu} for studying the DMT of the multiple-access channel (MAC). Interesting subsequent work has been carried out in \cite{ITW11-roopefransu,roopefransujournal,roopelaura_ISIT}. 
\end{remark}

Next, let us denote by 
$$
S_K(m)=\sum_{k=1}^{N_m}\frac{b_k'}{k^{s}}
$$
the truncated sum, where $b_k'= 0$ for those $k$ that do not appear as a norm for $x\in\OO_K$, $H(x)\leq m$.  An upper bound for the truncated sum is achieved  from the truncated Dedekind zeta function $\zeta_K(s,m)$.

\begin{prop}(Upper Bound)
Let $\OO_K$ be a PID. Then we have that
$$
S_K(m)\leq\max\{b_k'\,|\, k\leq N_m\}\cdot\zeta_K(s,m).
$$
\end{prop}
\begin{IEEEproof}
First, note that if $a_k=0$, then $b_k=0$ and hence $b_k'=0$ since $\OO_K$ is a PID. Now a simple computation gives us

\begin{eqnarray*}
&&S_K(s,m)=\sum_{k\leq N_m}\frac{b_k'}{k^{s}}\\
&\leq&\max\{b_k'\,|\, k\leq N_m\}\sum_{b_k'\neq 0,k\leq N_m}\frac 1{k^{s}}\\
&\leq&\max\{b_k'\,|\, k\leq N_m\}\cdot\zeta_K(s,m).\\
\end{eqnarray*}
In the second step, the summation is only taken over the values of $k$ for which $b_k'\neq 0$, \emph{i.e.}, $k$ appears as a norm for some element $x$. Hence, for all the terms $1/k^s$ included in the sum $a_k\neq 0$. In addition to this, the truncated zeta sum may contain other terms (for which $b_k\neq 0$ but $b_k'=0$ due to the energy limit) so we indeed get an upper bound. 
\end{IEEEproof}

\begin{remark}
We anticipate that the above bounds are not very tight, especially when extended to the non-PID case. Hence, our goal in the next section is to derive tighter bounds arising from geometric analysis. 
\end{remark}

\section{Probability bounds from geometric derivation}
\label{geometric}
Let us consider an algebraic lattice generated by the canonical embedding of the ring of integers $\OO_K$ of a totally real algebraic number field $K= {\QQ}(\theta)$ of degree $n$, 
\begin{eqnarray*}
\Lambda &=& \{ \xv = (x_1,\ldots ,x_n)= \\
&=&(\sigma_1(\alpha),\sigma_2(\alpha),\cdots, \sigma_k(\alpha)) \, |\, \alpha \in \OO_K \},
\end{eqnarray*}
where the $\sigma_i$'s are the $n$ field homomorphisms.
In the following we will indistinctly talk about lattice points and algebraic integers.

We will consider a finite constellation $\mathcal{S} = \Lambda \cap \Bc$ where the bounding region
is a hypercube of side $2R$ centered at the origin.

The algebraic norm of an algebraic integer is equal to the 
product distance of the corresponding lattice point from the origin,
\[
N(\alpha) =  \prod_{i=1}^n \sigma_i(\alpha) = \prod_{i=1}^n x_i\,.
\]
Since the field norms of algebraic integers to $\mathbf{Q}$ are integers
we observe that all the lattice points on the hyperbolic naps
\[
\prod_{i=1}^n |x_i| = k
\]
correspond to algebraic integers with the same absolute norm. 
When $k=1$ we have the group of units of $K$ which are fully characterized by the Dirichelet theorem.

In order to evaluate the union bound on the error probability and  the probability of the eavesdropper's correct decision, we now use a slightly different notation for the truncated sum as a product distance theta series, 
\begin{eqnarray} \label{eq_pe}
S_K(s,\mathcal{S}) = \sum_{\mathcal{S},\, k\geq 1} \frac{n_k}{k^s},
\end{eqnarray}
where $n_k$ is the number of constellation points at product distance $k$ from 
the origin when the sums are limited to the points in the finite constellation $\mathcal{S}$.

We need to count the number of lattice constellation points on each hyperbolic nap.
We will consider the following logarithmic coordinate system defined by 
\[
X_i = \log|x_i|,\hspace{5mm} i=1,\ldots,n.
\]
In this logarithmic space the hyperbolic naps become hyperplanes $\Hc_k$
perpendicular to the all ones vector. In particular, the units are on the 
hyperplane passing through the origin of the log-space.
We also note that the number of hyperbolic naps that are mapped to the same lattice is given by the number $w$ of roots of unity that are in the field $K$.

The bounding box $\Bc$ is transformed in the log-space to a semi-infinite rectangular region with a corner at
\[
(\log(R),\ldots, \log(R))
\]
denoted by $\log(\Bc)= (-\infty,\log(R)]^n$.

Since $k$ is positive, the intersection  
\[
\mathcal{S}_k=\log(\Bc) \cap \Hc_k 
\]
is nonempty for all absolute norms $1\leq k \leq R^n$.

Dirichelet's Theorem proves that the units of $K$ form a lattice $\Lambda_{\log}$ in the hyperplane
\[
\sum_{i=1}^n X_i= 0
\]
with a fundamental volume Vol$(\Lambda_{\log})=\rho_K\sqrt{r_1+r_2}$, where $\rho_K$ is the  regulator  of the number field (cf. Def. \ref{regulator}) and $n=r_1+2r_2$ ($n=r_1$ when $K$ totally real).
 
This suggests a simple and well-known way to estimate the number of units in the box as  
\[
n_1 = \left\lfloor \frac{w \vol(\mathcal{S}_1)}{ \vol(\Lambda_{\log})}  \right\rfloor,
\]
where $\lfloor\cdot\rfloor$ denotes the floor operation.

For non units ($k>1$) the problem is more complicated. We need to count the number $a_k$ of principal ideals $I_j(k)=(\alpha_j)$ of norm $k$ generated by the elements $\alpha_j$ of absolute norm $k$. Since $N(\alpha u) = N(\alpha)$ for all units $u$  and the
norm of a principal ideal is equal to the absolute norm of its generator we can conclude that
the set of points on the hyperplane $\mathcal{S}_k, k>1$ is a union of translates of the logarithmic  lattice. Then we can estimate the number of points of norm $k>1$ in the box as 
\[
n_k = \left\lfloor \frac{w a_k \vol(\mathcal{S}_k)}{ \vol(\Lambda_{\log})}  \right\rfloor.
\]

\begin{definition}
If we denote by $b_k$ the exact number of points of absolute norm $k$, then we define an \emph{error function} $$f_k(R)=\lfloor|n_k-b_k|\rfloor,$$ which is  dependent on the edge length $R$. 
\end{definition}

\begin{remark}
It is known that the error function grows quite large when the dimension of the lattice grows. We will illustrate the size of the error function later in Section \ref{examples}. We are not making any claims as to how tight our estimate $n_k$ is asymptotically, but rather wish to show that with small values of $n$ with practical interest it will give us a good estimate. 
\end{remark}

Let us now focus on the derivation of $\vol(\mathcal{S}_k)$. We first note that $\mathcal{S}_k$ is the basis
of a hyper pyramid $V_k$ with vertex in $(\log(R),\ldots, \log(R))$ whose volume is 
equal to the volume of a simplex with $n$ orthogonal vectors of length 
$n \log(R)- \log(k)$, i.e.,
\[
\vol(V_k) = \frac{(n \log(R)- \log(k))^n}{n!}\,.
\]
The height of $V_k$ is given by
\[
h_k = \frac{n \log(R)- \log(k)}{\sqrt{n}}\,.
\]
Hence
\begin{eqnarray*}
\vol(\mathcal{S}_k) &=& \frac{n \vol(V_k)}{h_k} \\
&=&  \frac{\sqrt{n}}{(n-1)!}(n \log(R)- \log(k))^{n-1}.
\end{eqnarray*}

We can finally write
\begin{eqnarray} \label{eq_pe2}
S_K(s,\mathcal{S}) &\leq&  \sum_{\mathcal{S},\, k\geq 1} \frac{n_k}{k^s}\\
& =& \sum_{\mathcal{S},\, k=1}^{\lfloor R^n \rfloor} \frac{w a_k \vol(\mathcal{S}_k)}{ \vol(\Lambda_{\log}) k^s} \\
&=&  \sum_{\mathcal{S},\, k=1}^{\lfloor R^n \rfloor} \frac{w a_k \sqrt{n}}{(n-1)! \vol(\Lambda_{\log}) k^s} \\
&&\cdot(\log(R^n)-\log(k))^{n-1}. 
\end{eqnarray}

Let us now develop this sum further by using the Dedekind zeta function.
The Dedekind zeta function $\zeta_K(s)$ is first written as a Dirichelet series defined by the coefficients $a_k$,
\[
 \zeta_K(s) = \sum_{k=1}^\infty \frac{a_k}{k^s}.
\]

Using the fact that 
\[
D_s^{(m)} \zeta_K(s) = (-1)^m\sum_{k=1}^\infty \frac{a_k(\log(k))^m}{k^s}
\]
and the binomial expansion 
\begin{eqnarray*}
&&(\log(R^n)-\log(k))^{n-1} \\
&=& \sum_{m=0}^{n-1} {{n-1} \choose {m}} 
(\log(R^n))^{n-1-m}(\log(k))^m
\end{eqnarray*}
we can write an upper bound to  $S_K(s,\mathcal{S})$ as a function of  
$\zeta_K(s)$ and its derivatives at $s$. We denote
$$
K_1=\frac{w\sqrt{n}}{(n-1)! \vol(\Lambda_{\log})}.
$$
Now\begin{footnotesize}
\begin{eqnarray*}
&S_K(s,\mathcal{S})&\leq K_1 \sum_{\mathcal{S},\, k=1}^{\lfloor R^n \rfloor} 
\frac{a_k}{ k^s} (\log(R^n)-\log(k))^{n-1} \\
&=&K_1\sum_{\mathcal{S},\, k=1}^{\lfloor R^n \rfloor}\left[ 
\frac{ a_k } {k^s} \sum_{m=0}^{n-1}{{n-1} \choose {m}}(\log(R^n))^{n-1-m}(\log(k))^{m}\right]\\
&=&K_1 \sum_{m=0}^{n-1}\left[{{n-1} \choose {m}}(\log(R^n))^{n-1-m}\sum_{\mathcal{S},\, k=1}^{\lfloor R^n \rfloor}\frac{ a_k(\log (k))^m } {k^s}\right].\label{dedekind}\\
\end{eqnarray*}
\end{footnotesize}
Next let us upper bound the latter sum in \eqref{dedekind} by letting the sum go to infinity:
\begin{eqnarray*}
\sum_{\mathcal{S},\,k = 1}^{\lfloor R^n \rfloor}\frac{ a_k(\log (k))^m } {k^s}
&\leq&\sum_{k = 1}^{\infty}\frac{ a_k(\log (k))^m } {k^s}\\
&=&|D_s^{(m)}\zeta_K(s)|.
\end{eqnarray*}

Inserting this to \eqref{dedekind}, we finally get
\begin{equation}
S_K(s,\mathcal{S})\leq  K_1\sum_{m=0}^{n-1}{{n-1} \choose {m}}(\log(R^n))^{n-1-m}|D_s^{(m)}\zeta_K(s)|.
\end{equation}
Concentrating on the leading term we further get
\begin{equation}
S_K(s,\mathcal{S})\leq  K_1(\log(R))^{n-1}|\zeta_K(s)|.
\end{equation}

Accurate numerical functions enable to easily evaluate this bound. For instance, one can use the Taylor expansion of the Dedekind zeta function implemented in Sage in order to calculate the derivatives. Note that the sum only contains $n$ terms, which indeed makes the computation feasible.

\section{Examples of the accuracy of $n_k$} 
\label{examples}

\begin{figure}[t]
\begin{center}
\includegraphics[width=13cm]{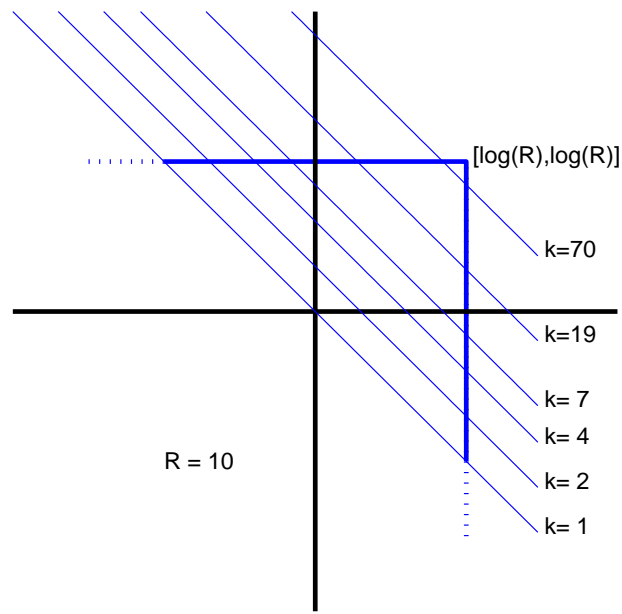}
\end{center}
\caption{Illustration of the logarithmic lattice with $n=2$ and $R=10$. }
\label{Fig2D}
\end{figure}

\begin{figure}[t]
\begin{center}
\includegraphics[width=13cm]{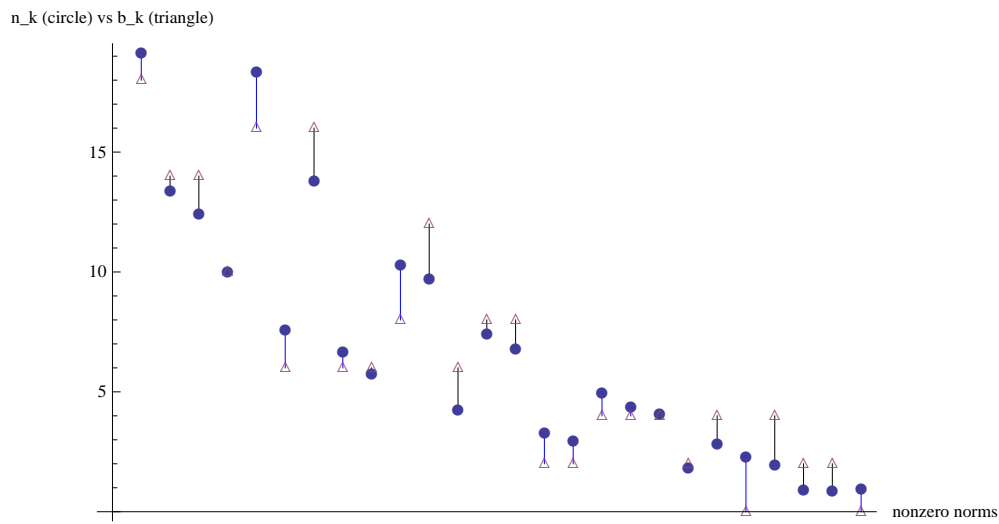}
\end{center}
\caption{Illustration of the estimation error,  $n=2$ and $R=10$.}
\label{Fign2pieni}
\end{figure}

\begin{figure}[t]
\begin{center}
\includegraphics[width=13cm]{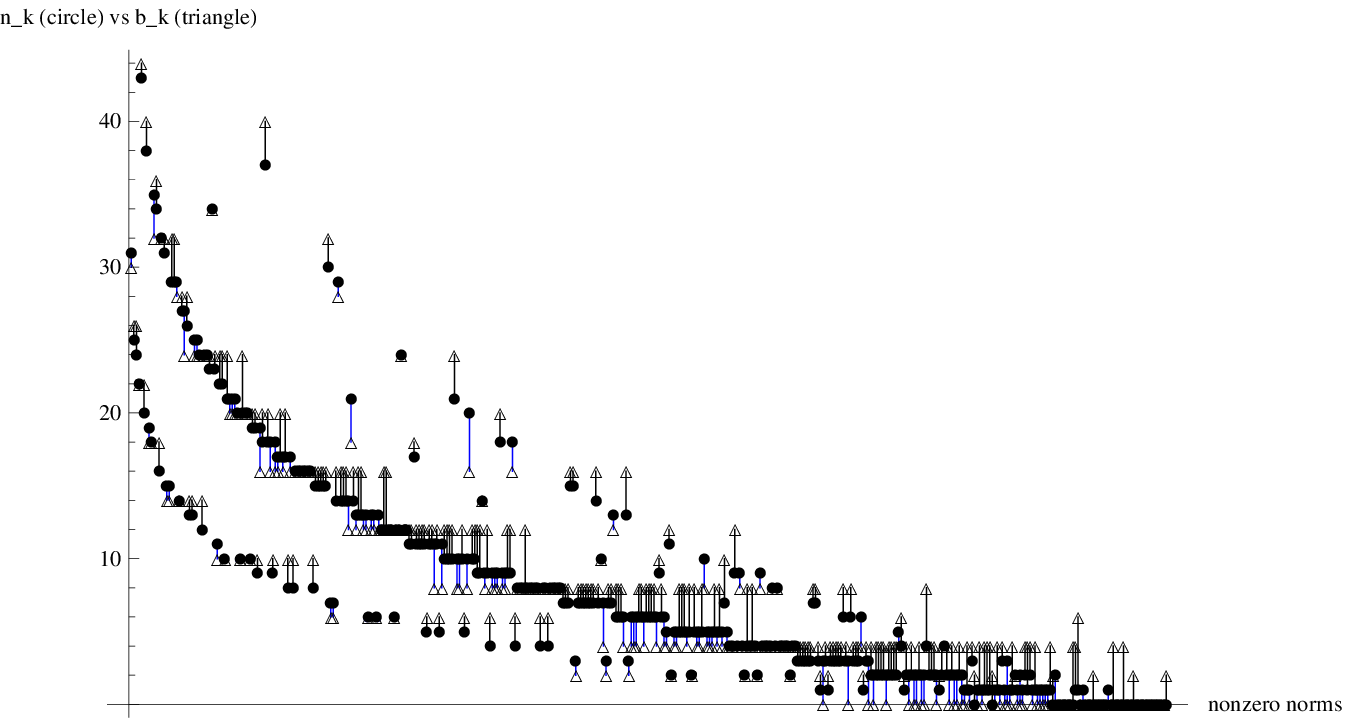}
\end{center}
\caption{Illustration of the estimation error, $n=2$ and $k\geq 2000$. }
\label{Fign2iso}
\end{figure}

\begin{figure}[t]
\begin{center}
\includegraphics[width=13cm]{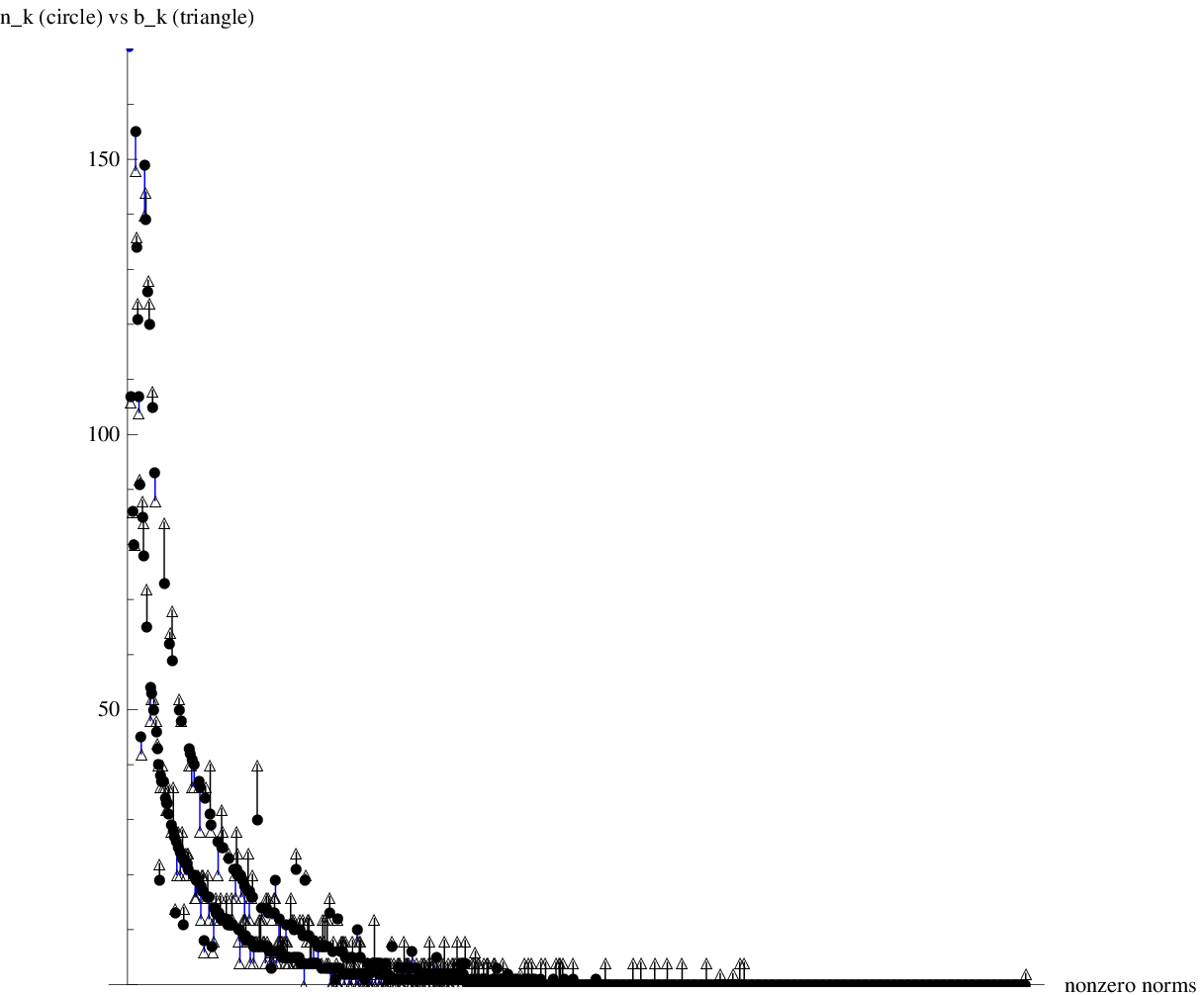}
\end{center}
\caption{Illustration of the estimation error,  $n=4$ and $k\geq 10000$. }
\label{Fign4}
\end{figure}

\begin{figure}[ht]
\begin{center}
\includegraphics[width=13cm]{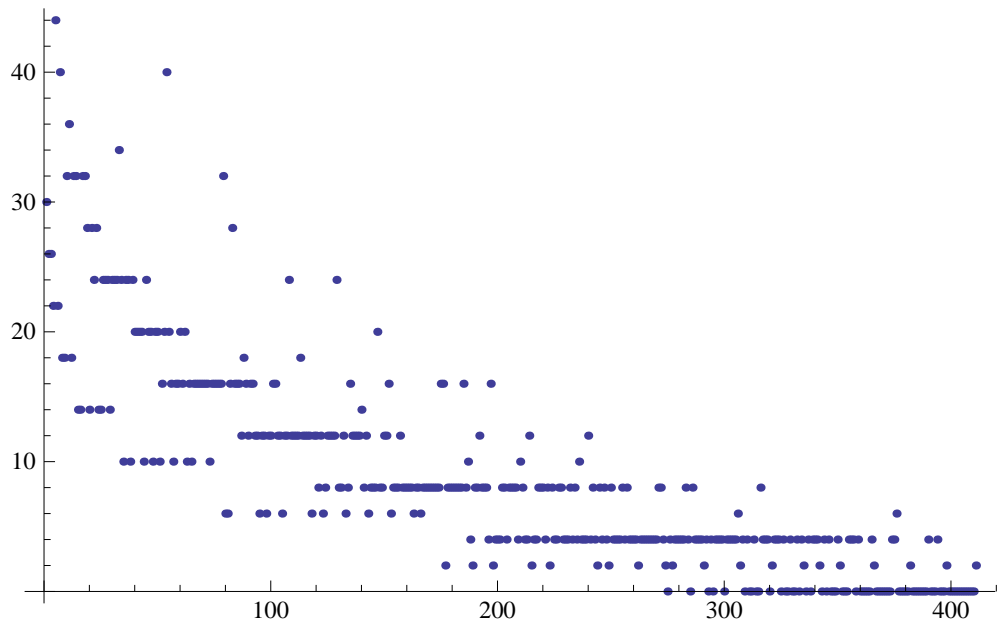}
\end{center}
\caption{The exact values take a staricase form.}
\label{Figstair}
\end{figure}

\begin{figure}[ht]
\begin{center}
\includegraphics[width=13cm]{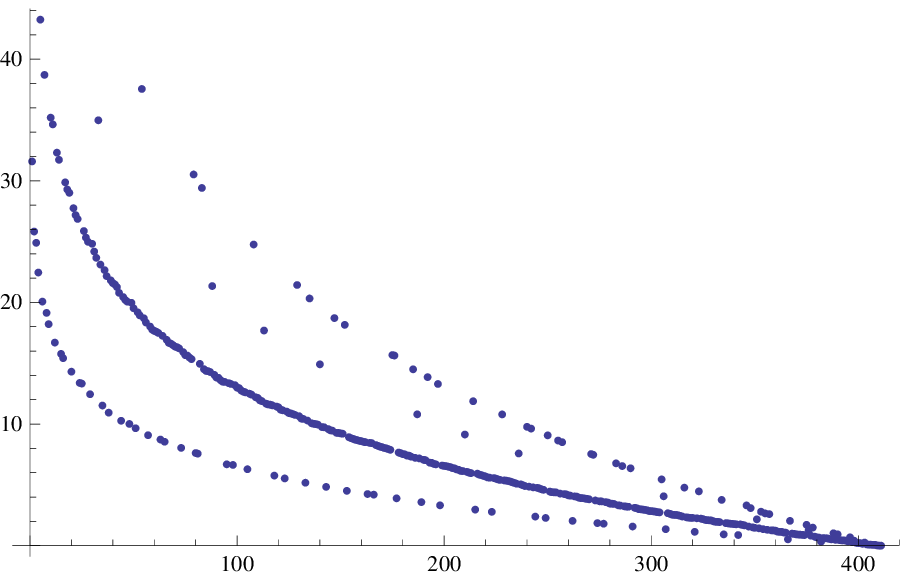}
\end{center}
\caption{The estimates form a smooth function.}
\label{Figsmooth}
\end{figure}

Let us consider the field extension $K/\Q$  with $K=\mathbf{Q}(\sqrt{5})$ of degree $n=2$. As $K$ is totally real, we  have  $w=2$.  The regulator of the field is
$
\rho_K=0.481211825059603.
$

Let us first set $R=10$ (see Figure 2), i.e., $1\leq k\leq 100$, and denote by $b_k$ the exact number of lattice points of absolute norm $k$ within the square centered at the origin with the edge length $2R$, i.e.,
$$
b_k=\#\{\mathbf{x}\in \Lambda\cap \mathcal{B}\, |\, N_{F/\Q}(x)=k\}.
$$
The values of $n_k$ (without the floor operation), $b_k$, and $f_k(R)=\lfloor |n_k-b_k| \rfloor$ (the line connecting the previous two) are collected in Figure 2. We can see that the error $$f_k(R)=\lfloor|n_k-b_k|\rfloor \leq 2$$ for all $k$. The values are only given for those $k$ for which $a_k\neq 0$, that is, there exists a principal ideal of norm $k$. For all other $k$ we have $n_k=b_k=f_k(R)=0$.

When we increase the size of the constellation by considering norms up to $k=2000$, i.e., $R=2000$, we still have  
$$f_k(R)=\lfloor|n_k-b_k|\rfloor \leq 3\quad \forall k,$$
see Figure 3.

In Figure 4 we depict the same plot for a lattice $\Q(\alpha)/\Q$ with dimension $n=4$ and minimal polynomial $f_\alpha(x)=x^4-x^3-3x^3+x+1$ and for norms up to $k=10000$. As we can see, the points are well aligned and the error remains small,
$$f_k(R)=\lfloor|n_k-b_k|\rfloor \leq 3\quad \forall k.$$ 

\begin{figure}[h!]
\begin{center}
\includegraphics[width=13cm]{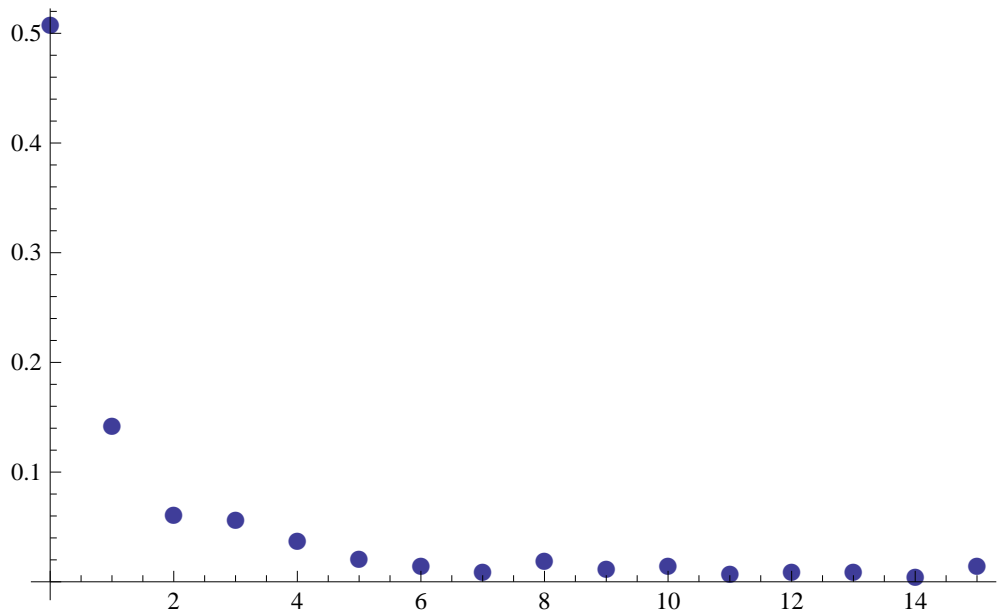} 
\end{center}
\caption{The frequency of estimation errors for $k$, $1\leq k \leq 65536$. Edge length $=2R=10$ and $n=8$. }
\label{n8error1}
\end{figure}

\begin{figure}[h!]
\begin{center}
 \includegraphics[width=13cm]{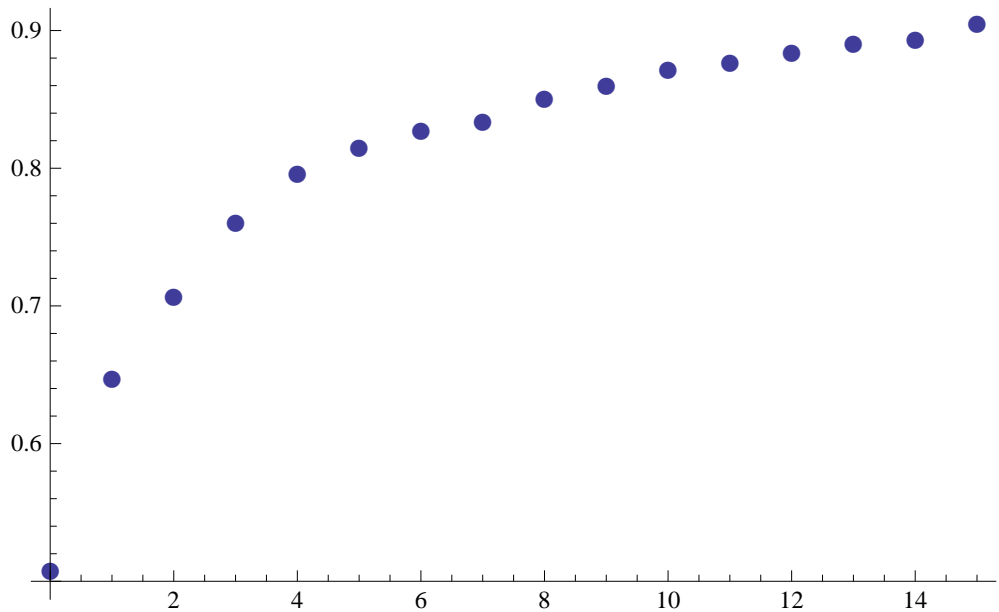}
\end{center}
\caption{The  cumulative frequency  of estimation errors for $k$, $1\leq k \leq 65536$. Edge length $=2R=10$ and $n=8$. }
\label{n8error}
\end{figure}

In order to see what happens to the size of error in the estimate $n_k$ when the dimension grows, let us consider a case with $n=8,\, w=2$. This is already quite a high delay in practice, as we require encoding over eight time instances. 

The construction is based on the maximal totally real subfield of the 32nd cyclotomic field, namely  $$K=\Q(\zeta_{32}+\zeta_{32}^{-1})$$
having a regulator $\rho_K=28.4375954169998$. 

While the absolute error gets bigger when dimension grows, it is still negligible considering that out of all the cases more than half correspond exactly to the estimate (i.e., $n_k=b_k$, meaning no error), and in the rest of the cases (meaning an error occurs) either the error is very small or (a bigger error) occurs very rarely.  In Figures \ref{n8error1} and \ref{n8error} we have depicted the frequency and cumulative frequency of errors, respectively. One can see that  cumulative frequency as high as 90\% is achieved already by errors of size  $\leq 15$.

\begin{remark} It is worth pointing out that the error induced by the estimate is partly due to the fact that we are essentially trying to estimate a staircase function by a smooth function, see Figures 5 and 6. 
\end{remark}

\begin{figure}[t]
\begin{center}
\includegraphics[width=13cm]{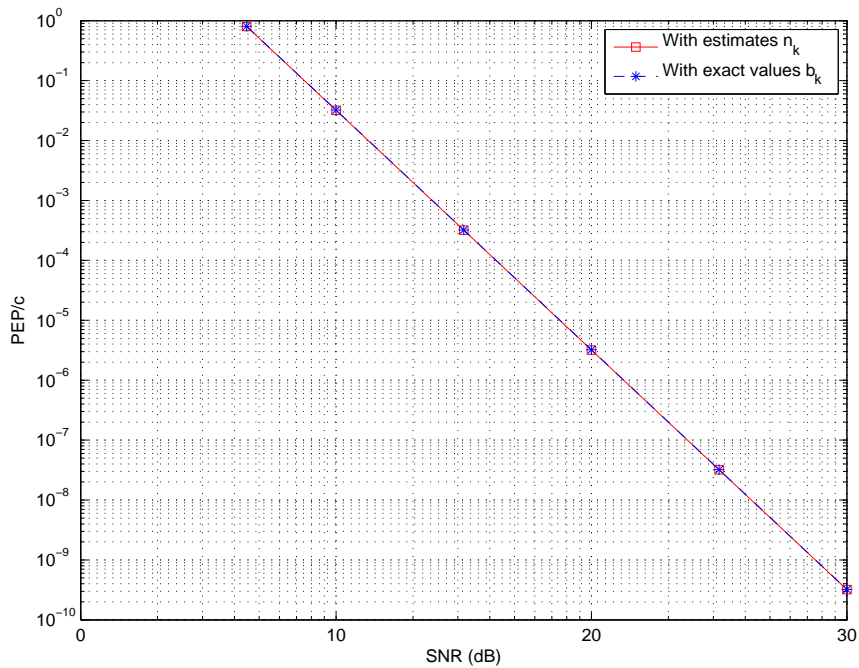}
\end{center}
\caption{PEP/$c$ as a function of SNR using 1) the estimate $n_k$ and 2) the exact values $b_k$. Extension $\Q(\alpha)/\Q$ with $f_\alpha(x)=x^4-x^3-3x^3+x+1$, $n = 4,\, R = 10,\,1\leq k\leq  10000$.}
\label{wiretap_snr}
\end{figure}

Finally, we plot the theoretical estimated and exact PEP curves $P_e(SNR)$ as a function of SNR. Figure
 \ref{wiretap_snr} shows that there is no penalty by using the estimates $n_k$ in place of the exact values $b_k$ when computing \eqref{pep}; the curves are perfectly aligned. We have ignored the constant $c$ since this would be the same for both sums, and simply plotted $P_e=\frac{1}{\gamma^n}\sum_{k=1}^{10000}\frac{n_k}{k^2}$ vs $P_e=\frac{1}{\gamma^n}\sum_{k=1}^{10000}\frac{b_k}{k^2}$, letting $\gamma$ take values over an SNR range.

\section{Conclusions and future work}\label{conclusions}
In this paper, we have considered number field lattice codes and  provided probability  bounds for the PEP and for the probability of a correct decision of the eavesdropper on a fast fading channel using Dedekind zeta functinos. More refined bounds were then derived using geometric analysis on logarithmic lattices. As a byproduct, an estimate for the number of constellation points with certain algebraic norm was given, and its accuracy was demonstrated through practical examples.

Future work will consist of improved analysis on the error made in the estimation and of generalizing the results  to complex lattices and MIMO channels. One promising approach is offered by division algebras, along the same lines as in  \cite{roopefransujournal,roopelaura_ISIT}.

\section{Acknowledgments} The authors would like to thank Prof. Fr\'ed\'erique Oggier and Dr. Roope Vehkalahti for useful discussions. 







\end{document}